# From Labyrinthine Patterns to Spiral Turbulence

Aric Hagberg
*Program in Applied Mathematics*
*University of Arizona, Tucson, AZ 85721*

Ehud Meron
*Arizona Center for Mathematical Sciences*
*Department of Mathematics*
*University of Arizona, Tucson, AZ 85721*

A new mechanism for spiral vortex nucleation in nongradient reaction diffusion systems is proposed. It involves two key ingredients: An Ising-Bloch type front bifurcation and an instability of a planar front to transverse perturbations. Vortex nucleation by this mechanism plays an important role in inducing a transition from labyrinthine patterns to spiral turbulence.

05.45.+b, 82.20.Mj

Labyrinthine patterns provide a beautiful example of pattern formation phenomena near equilibrium. Patterns of this form have been observed in various gradient systems including garnet layers [1], ferrofluids [2] and block copolymers [3]. Recently, Lee *et al.* found labyrinthine patterns in a bistable chemical reaction maintained far from equilibrium [4]. Unlike equilibrium systems, nongradient systems of this kind may undergo nonequilibrium Ising-Bloch transitions [5-7]. Such transitions or bifurcations render stationary (Ising) fronts unstable and give rise to pairs of counter-propagating (Bloch) fronts. In this Letter we show the approach to an Ising-Bloch bifurcation may induce a transition from labyrinthine patterns to "spiral turbulence", that is, a state of spatiotemporal disorder with repeated events of spiral vortex nucleation and annihilation.

Spiral turbulence has been found recently in reaction-diffusion models of cardiac tissues [8-10] and surface reactions [11]. In both contexts wavetrain instabilities were found to precede the break up of spiral waves into new pairs of spiral waves. In this sense the resulting state of spiral turbulence is analogous to defect mediated turbulence in dissipative systems undergoing phase instabilities [12]. Here we report a different mechanism for spiral vortex nucleation that involves two generic front instabilities: An Ising-Bloch front bifurcation, and an instability to transverse perturbations [13,14]. The front bifurcation provides the system with two coexisting Bloch fronts. The transverse instability provides the driving force needed to convert a segment of a Bloch front propagating in one direction into a Bloch front propagating in the opposite direction. This process nucleates a pair of spiral vortices, one at each end of the segment. Such nucleation events can be realized by approaching the front bifurcation point from the regime of labyrinthine patterns, for the latter already imply the existence of a transverse instability.

We study spiral vortex nucleation using a nongradient reaction-diffusion model of the FitzHugh-Nagumo type:

$$u_t = u - u^3 - v + \nabla^2 u, \qquad (1a)$$

$$v_t = \epsilon(u - a_1 v - a_0) + \delta \nabla^2 v, \qquad (1b)$$

where $u$ and $v$ are two scalar real fields. The model contains four parameters: $\epsilon > 0$, the ratio between the time scales associated with the two fields; $\delta \geq 0$, the ratio between the two diffusion constants; and the two parameters, $a_1 > 0$ and $a_0$. We choose $a_0$ and $a_1$ so that the system (1) has two stable uniform steady states: A "down" state $(u_-, v_-)$ and an "up" state $(u_+, v_+)$. Note that for $a_0 = 0$ the up and down states have the symmetry $(u_-, v_-) = -(u_+, v_+)$. This model has the interesting property that for $\epsilon \gg 1$ it can be reduced to a nonlocal gradient system by eliminating $v$ adiabatically [15]. Similar nonlocal models have been studied in the context of phase separating systems [16,17]. The regime $\epsilon \gg 1$ has been studied most recently by Petrich and Goldstein who derived heuristically a nonlocal equation of motion for the front interface separating the up and down states [18].

The Ising-Bloch front bifurcation for the system (1) has been studied in Refs. [7,19]. In the regime $\delta/\epsilon \gg 1$ and for $a_1$ sufficiently large, the speed $c$ of a planar front connecting the up state at $x = -\infty$ to the down state at $x = \infty$ satisfies

$$c = \frac{3c}{\sqrt{2}q^2(c^2 + 4\eta^2 q^2)^{1/2}} + c_\infty, \qquad (2)$$

where $\eta^2 = \epsilon\delta$, $q^2 = a_1 + 1/2$ and $c_\infty = \frac{3a_0}{\sqrt{2}q^2}$. In the symmetric case ($a_0 = 0$) a graph of $c = c(\eta)$ yields a pitchfork bifurcation diagram. For $\eta > \eta_c = \frac{3}{2\sqrt{2}q^3}$, $c = 0$ is the only solution, representing a stationary Ising front. For $\eta < \eta_c$, two additional solution branches appear, $c = \pm 2q\sqrt{(\eta_c^2 - \eta^2)}$, representing counter propagating Bloch fronts. Structurally, a Bloch front differs from an Ising front in that the $v$ field is displaced with respect to the $u$ field. The magnitude and the sign of the displacement determine the speed and the direction of propagation, respectively. A useful rule to remember is that the $v$ field always lags behind the $u$ field.

A typical form of the bifurcation diagram for the nonsymmetric case is shown in Fig. 1. Front multiplicity arises through a saddle-node bifurcation at a critical $\eta$ value, $\eta_c(a_0)$. We will refer to the single front solution that exists for $\eta > \eta_c$ as an Ising front and to the two stable front solutions for $\eta < \eta_c$ as Bloch fronts. With this convention the Ising front and one of the Bloch fronts lie on the same solution branch. Note that the Ising front is not stationary.

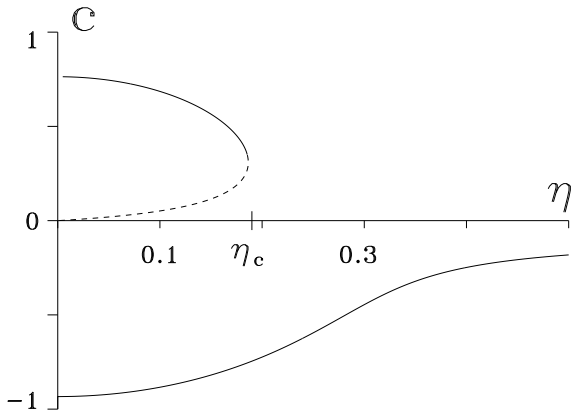

FIG. 1. A front bifurcation diagram for the nonsymmetric case, obtained from equation (2). Parameters used: $a_0 = -0.1$, $a_1 = 2$.

To study the transverse instabilities of the various front solutions we consider a weakly curved front. Introducing an orthogonal coordinate system $(s, r)$ moving with the front, the equations (1) assume the leading order form

$$u_{rr} + (c_r + \kappa)u_r + u - u^3 - v = 0, \quad (3a)$$

$$\delta v_{rr} + (c_r + \delta\kappa)v_r + \epsilon(u - a_1 v - a_0) = 0. \quad (3b)$$

The coordinate $s$ parametrizes the front and $c_r(s,t)$ and $\kappa(s,t)$ are, respectively, its normal speed and curvature. Both $c_r$ and $\kappa$ may vary weakly along the front and evolve slowly in time. Multiplying (3b) by the factor $\Delta(s,t) = (c_r + \kappa)/(c_r + \delta\kappa)$ the resulting system can be interpreted as equations for a planar front propagating at speed $c_r + \kappa$ along the $r$ axis in an effective medium characterized by the parameters $\tilde{\epsilon} = \epsilon\Delta$ and $\tilde{\delta} = \delta\Delta$ [20]. Using equation (2) with $c$ replaced by $c_r + \kappa$ and $\eta$ by $\tilde{\eta} = \eta\Delta$ we obtain

$$c_r + \kappa = \frac{3(c_r + \delta\kappa)}{\sqrt{2}q^2[(c_r + \delta\kappa)^2 + 4\eta^2 q^2]^{1/2}} + c_\infty. \quad (4)$$

Equation (4) can be solved for $c_r$ in terms of $\kappa$. Since $\kappa$ is small we expand $c_r = c_0 + d\kappa + \mathcal{O}(\kappa^2)$, where $c_0(\eta)$ is the speed of a planar front satisfying (2), and evaluate the prefactor $d$. This yields

$$d = \frac{1}{\alpha} + (1 - \frac{1}{\alpha})\delta, \quad \alpha = 1 - \frac{c_0 - c_\infty}{c_0}\left[1 - \frac{2q^4}{9}(c_0 - c_\infty)^2\right]. \quad (5)$$

Setting $d = 0$ for a given branch of $c_0$ gives the parameter values at which the corresponding planar front becomes unstable to transverse modulations. For the symmetric case ($a_0 = 0$) the Ising and Bloch fronts become unstable to transverse modulations when $\delta > \delta_I(\epsilon) = \frac{8}{9}q^6\epsilon$ and $\delta > \delta_B(\epsilon) = \frac{3}{2\sqrt{2}q^3\sqrt{\epsilon}}$, respectively. The transverse instability lines for the nonsymmetric case are shown in Fig. 2. Notice that the lines corresponding to the two Bloch fronts, denoted by $\delta_B^\pm$, are not degenerate as in the symmetric case. Also shown in this figure is the front bifurcation line, $\eta = \eta_c$ which we write as $\delta = \delta_F(\epsilon)$. For the symmetric case $\delta_F(\epsilon) = \frac{9}{8q^6\epsilon}$.

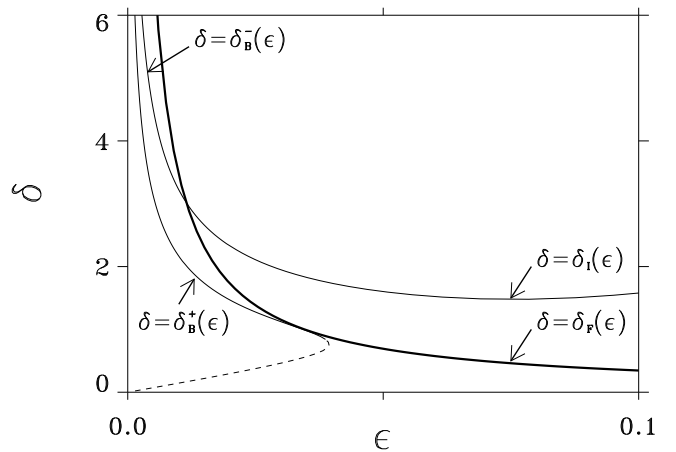

FIG. 2. The front bifurcation line (thick) and the transverse instability lines (solid thin) for the nonsymmetric case, obtained from equations (2) and (5) respectively. The dashed line represents the transverse instability of the unstable front branch (dashed line in Fig. 1). Parameters used: $a_0 = -0.1$, $a_1 = 2$.

The front bifurcation line and the transverse instability lines divide the $\epsilon - \delta$ parameter plane into regions where different pattern behaviors are expected. In the Ising regime, $\delta > \delta_F(\epsilon)$, and below the transverse instability line of Ising fronts, $\delta < \delta_I(\epsilon)$, we find stationary stripe patterns or, for $\delta$ sufficiently small, no patterns at all. In the Bloch regime, and below the transverse instability lines, $\delta = \delta_B^\pm$, traveling waves, including rotating spiral waves, prevail [7]. We emphasize though that the bifurcation and the transverse instability lines were derived for planar fronts and that curvature may shift these lines. In particular, it is possible to have two appropriately curved fronts propagating in opposite directions on the Ising side of the front bifurcation line.

Fig. 3 shows the time evolution of a stripe of up state domain well inside the Ising regime and *above* the transverse instability line, obtained by numerical integration of (1). The stripe is transversally perturbed in its middle part and at first a meandering stripe forms. At later stages, fingering and tip splitting processes take place until a stationary labyrinthine pattern fills up the whole system. An important characteristic of the final pattern is that the up-state domain is connected; domain fission is avoided because of repulsive front interactions. This behavior is very similar to that observed by Lee *et al.* in the bistable chemical recation [4] and to the simulations of Petrich and Goldstein [18] on their nonlocal interface model. We note that the value of $\epsilon$ used in producing Fig. 3 was quite small ($\epsilon = 0.05$). This suggests the gradient nature of (1) for $\epsilon \gg 1$ persists for smaller $\epsilon$ values as well, extending the validity range of the nonlocal interface model.

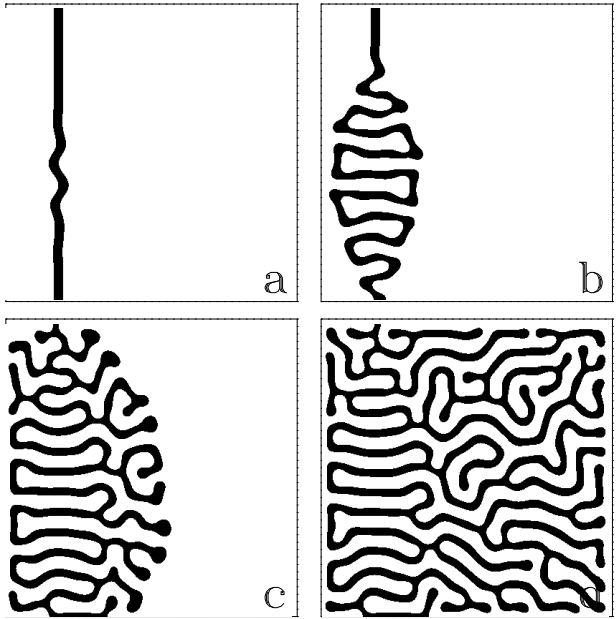

FIG. 3. The development of a labyrinthine pattern from a single stripe in the Ising front regime. The light and dark regions correspond to the down and up states, respectively. The frames a,b,c,d pertain to times $t = 100, 950, 1900, 5000$. Parameters used: $a_0 = -0.1$, $a_1 = 2$, $\epsilon = 0.05$, $\delta = 4$.

Sufficiently close to the front bifurcation line, however, nongradient effects become important. Fig. 4 shows the time evolution of a stripe in the vicinity of that line. With this parameter choice both the leading and the trailing fronts of the stripe (Bloch fronts of different kinds) are unstable to transverse perturbations, but the instability of the leading front is stronger. In contrast to the time evolution shown in Fig. 3 the stripe breaks into disjoint pieces which subsequently develop into a complex spatiotemporal pattern. We have continued the simulation up to $t = 5000$ without finding any qualitative change. The number of vortices appears to fluctuate between 30 to 70. We point out, however, that equations (1) always have the two simple attractors, $(u_-, v_-)$ and $(u_+, v_+)$, and, in principle, further evolution might culminate in one of them.

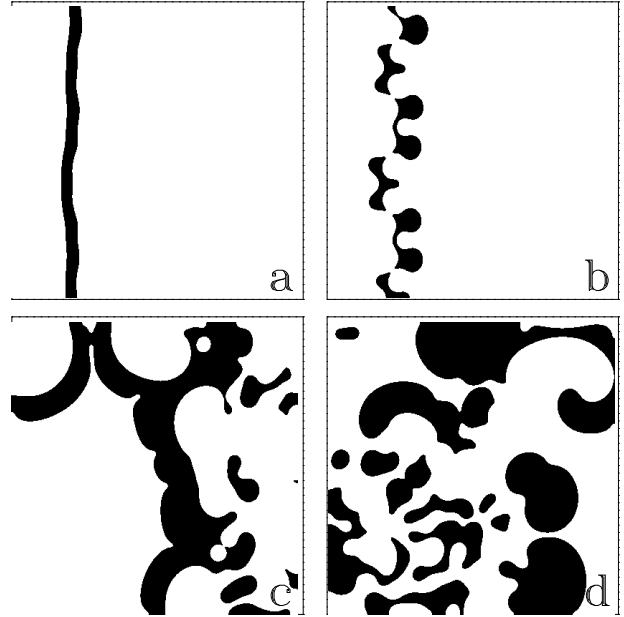

FIG. 4. The development of spiral turbulence from a single stripe close to the front bifurcation line. The light and dark regions correspond, respectively, to the down and up states. The frames a,b,c,d pertain to times $t = 30, 210, 1180, 1800$. Parameters used: $a_0 = -0.1$, $a_1 = 2$, $\epsilon = 0.014$, $\delta = 2.8$.

Fig. 5 provides a closer look at the processes involved in Fig. 4. Shown in this figure are the contours $u = 0$ (thick line) and $v = 0$ (thin line) for four consecutive time frames. The frames show a dent in a Bloch front propagating to the right that turns into a Bloch front propagating to the left. This process is accompanied by the formation of a vortex pair appearing as crossing points of the two zero contours. The growing dent eventually cuts the up-state domain into two disjoint pieces in a process involving down-state domain fission and front reconnections. Domain fission is made possible at this parameter regime because of higher front speeds.

The rate of vortex nucleation depends not only on the strength of the transverse instability but also on the occasional formation of highly curved front regions. Such regions are most often created by the fission of two up-state domains, yielding cusp like structures as shown in Fig. 5a. This geometry facilitates the diffusive accumulation of $v$ at a small region, thereby inducing a local front transition and consequently vortex nucleation.

The simulations shown in Figs. 3 to 5 were obtained by integrating the system (1) using an implicit finite difference scheme for the labyrinthine patterns and an explicit scheme for the turbulent patterns. A $400 \times 400$ grid has been used with one grid point per unit length. This resulted in about 6 grid points across the front. Doubling the resolution slightly changed the location of the front bifurcation line but yielded the same qualitative results.

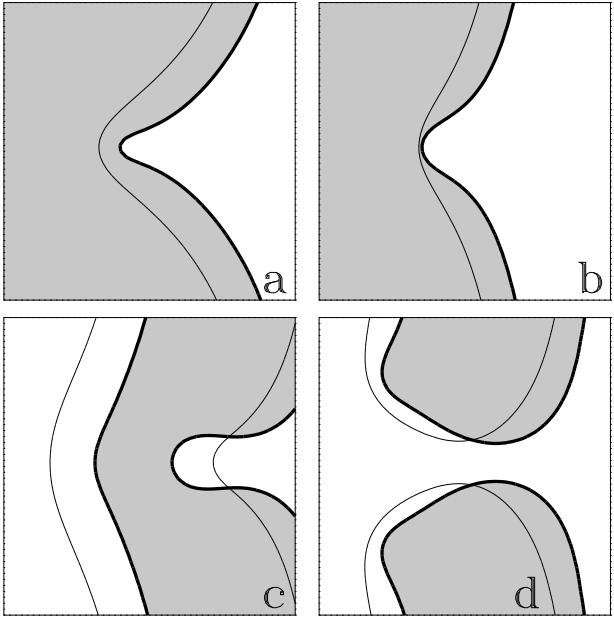

FIG. 5. Vortex nucleation, developing from a cusp like front shape, followed by domain fission and front reconnections. The thick (thin) line is a $u=0$ ($v=0$) contour. The light regions correspond to the down state and the shaded regions to the up state. The frames a,b,c,d pertain to times $t = 310, 390, 460, 500$. Parameters are as in Fig. 4.

We have shown here that the coupling between an Ising-Bloch bifurcation and a transverse front instability in a doubly-diffusive FitzHugh-Nagumo model may lead to spontaneous nucleation of spiral vortices. These nucleation events, together with the capability of domains to merge, can induce a transition to spiral turbulence. Similar behavior is expected to be found in other reaction diffusion systems exhibiting these two front instabilities. A possible candidate is the bistable reaction studied by Lee *et al.* [4].

We benefited from a talk given by Sergio Rica at the Nice workshop "From Excitability to Oscillations", June 1993. We also thank Alan Newell for interesting discussions. This work was supported in part by the Arizona Center for Mathematical Sciences (ACMS). ACMS is sponsored by AFOSR contract FQ8671-9000589 (AFOSR-90-0021) with the University Research Initiative Program at the University of Arizona. One of us (A.H.) acknowledges the support of the Computational Science Graduate Fellowship Program of the Office of Scientific Computing in the Department of Energy.

[1] M. Seul, L. R. Monar, L. O'Gorman, and R. Wolfe, Science **254**,1616 (1991).
[2] R. E. Rosensweig, *Ferrohydrodynamics* (Cambridge University Press, Cambridge, 1985).
[3] C. S. Henkee, E. L. Thomas, and L. J. Fetters, J. Mater. Sci **23**, 1685 (1988). See also T. A. Witten, Physics Today **43**, 21 (1990).
[4] K. J. Lee, W. D. McCormick, Q. Ouyang, and H. L. Swinney, Science **261**, 192 (1993).
[5] P. Coullet, J. Lega, B. Houchmanzadeh, and J. Lajzerowicz, Phys. Rev. Lett. **65**, 1352 (1990).
[6] A. Hagberg and E. Meron, Phys. Rev. E **48**, 705 (1993).
[7] A. Hagberg and E. Meron, "Pattern Formation in Non-Gradient Reaction-Diffusion Systems: The Effects of Front Bifurcations", Preprint (1993).
[8] M. Courtemanche and A. T. Winfree, Int. J. Bifurcation and Chaos **1**, 431 (1991).
[9] A. V. Holden and A. V. Panfilov, Int. J. Bifurcation and Chaos **1**, 219 (1991).
[10] A. Karma, Phys. Rev. Lett. **71**, 1103 (1993). See also the discrete models proposed by M. Gerhardt, H. Schuster, and J. J. Tyson, Science **247**, 1563 (1990); H. Ito and L. Glass, Phys. Rev. Lett. **66**, 671 (1991).
[11] M. Bär and M. Eiswirth, Phys. Rev. E **48**, R1635 (1993).
[12] P. Coullet, L. Gil, and J. Lega, Phys. Rev. Lett. **62**, 1619 (1989).
[13] T. Ohta, M. Mimura, and R. Kobayashi, Physica D **34**, 115 (1989).
[14] D. A. Kessler and H. Levine, Phys. Rev. A **41**, 5418 (1990).
[15] T. Ohta, A. Ito, and A. Tetsuka, Phys. Rev. A **42**, 3225 (1990).
[16] T. Ohta and K. Kawasaki, Macromolecules **19**, 2621 (1986).
[17] C. Roland and R. C. Desai, Phys. Rev. B **42**, 6658 (1990).
[18] D. M. Petrich and R. E. Goldstein, "A Nonlocal Contour Dynamics Model for Chemical Front Motion", Preprint (1993).
[19] H. Ikeda, M. Mimura, and Y. Nishiura, Nonl. Anal. TMA **13**, 507 (1989).
[20] A. S. Mikhailov and V. S. Zykov, Physica D **52**, 379 (1991).